\documentclass[aps,superscriptaddress,eqsecnum,nofootinbib,preprintnumbers]{revtex4}
\usepackage{graphicx,epsfig}
\usepackage{amsmath}
\usepackage {amssymb}
\usepackage{subfigure}

\usepackage{relsize}

\newcommand{\nn}{\nonumber}

\begin{document}

\title{

Avoidance of singularities in asymptotically safe Quantum Einstein Gravity
}

\author{Georgios Kofinas}
\email{gkofinas@aegean.gr} \affiliation{Research Group of Geometry,
Dynamical Systems and Cosmology,
Department of Information and Communication Systems Engineering\\
University of the Aegean, Karlovassi 83200, Samos, Greece}

\author{Vasilios Zarikas}
\email{vzarikas@teilam.gr} \affiliation{Department of Electrical Engineering, Theory Division\\
ATEI of Central Greece, 35100 Lamia, Greece}
\affiliation{Department of Physics, Aristotle University of Thessaloniki, 54124 Thessaloniki, Greece}

\begin{abstract}

New general spherically symmetric solutions have been derived with a cosmological ``constant''
$\Lambda$ as a source. This $\Lambda$ field is not constant but it satisfies the properties of the
asymptotically safe gravity at the ultraviolet fixed point. The importance of these solutions comes
from the fact that they describe the near to the centre region of black hole spacetimes
as this is modified by the Renormalization Group scaling behaviour of the fields. The consistent
set of field equations which respect the Bianchi identities is derived and solved. One of the
solutions (with conventional sign of temporal-radial metric components) is timelike geodesically complete,
and although there is still a curvature divergent origin, this is never approachable by an infalling
massive particle which is reflected at a finite distance due to the repulsive origin. Another family
of solutions (of both signatures) range from a finite radius outwards, they cannot be extended to
the centre of spherical symmetry, and the curvature invariants are finite at the minimum radius.

\end{abstract}

\maketitle

\section{Introduction}
\label{Introduction}

The hope that spacetime singularities in classical General Relativity are just an artifact of the high
degree of symmetries assumed in the solutions was soon lost when the singularity theorems of Hawking
and Penrose appeared. It is commonly believed that the appearance of spacetime singularities is an
indication of the limit of validity of General Relativity. A crucial aspect of any successful attempt
to quantise gravity is the ability to resolve somehow the break down of physics in these singular
regions of spacetime.

Although a considerable effort has been devoted to the discovery of a consistent fundamental theory of
nature that quantizes gravity, still this quest is one of the most challenging open problems in Science.
The perturbative quantization of General Relativity leads to a non-renormalizable quantum theory
\cite{tHooft:1974bx,Goroff:1985sz,vandeVen:1991gw}. Loop Quantum Gravity and spin foam models
assume that the spacetime is intrinsically discrete which offer an advance to the problem of
singularities. Concerning models on continuous spacetimes, one line of thought like String theory
suggests to search for new symmetries and new degrees of freedom in order to succeed an Ultra Violet
(UV) completion of General Relativity. Another research path is to keep the same fields and
symmetries from General Relativity and work on the belief that the associated gravitational action
is a fundamental part of the complete theory at the non-perturbative level. This is the idea behind
the Asymptotic Safety scenario \cite{livrev,oliverbook,reviews} which was first proposed by
Weinberg \cite{wein,Weinproc1}. The key issue is the existence of a non-Gaussian fixed point (NGFP)
of the renormalization group (RG) flow for gravity. If this NGFP determines the behavior of the theory
at the UV it can be proved that all measured quantities are free from nonphysical divergences.

The mathematical technique that gave boost to the asymptotic safety scenario is the functional
renormalization group equation for gravity \cite{mr} which enabled the detailed analysis
of the gravitational RG flow at the non-perturbative level
\cite{mr,percadou,oliver1,frank1,oliver2,oliver3,oliver4,souma,frank2,prop,perper1,
codello,litimgrav,essential,r6,MS1,Codello:2008vh,creh1,creh2,creh3,JE1,HD1,HD2,JEUM,
Rahmede:2011zz,Benedetti:2010nr}.
This technique uses a Wilsonian RG flow on a space of all theories that consists of all
diffeomorphism invariant functionals of the metric $g_{\mu \nu}$. The model emerging from this
construction is called Quantum Einstein Gravity (QEG). QEG does not emerge from a direct
quantization of classical General Relativity; its bare action corresponds to a non-trivial
fixed point of the RG flow and is a prediction assuming asymptotic safety.
In the asymptotic safety program one works with the gravitational effective average action $\Gamma_k$
which includes only the effect of the quantum fluctuations with momenta $p^2 > k^2$,
thus $\Gamma_k$ expresses an approximate description of physics at the momentum scale
$p^2 \approx k^2$. This effective description helps the development of the phenomenological
studies of asymptotic safety.

The phenomenology of the asymptotic safety proposal concerning the study of RG improved
cosmologies has been first appeared in \cite{Bonanno:2001xi,Bonanno:2001hi} (for a review
see \cite{Reuter:2012xf}).
The first RG improved solutions of black holes arising from asymptotic safety scenario were
derived in \cite{Bonanno:1998ye,Bonanno:2000ep}. Consequently, other
works extended this idea to the Vaidya metric \cite{Bonanno:2006eu}, while the modified Kerr
metric has been presented in \cite{Reuter:2006rg,Reuter:2010xb}. An analysis of the thermodynamic
properties of these black holes can be found in \cite{Falls:2012nd}, while a study of the possibility
to include higher derivative terms in the effective average action has been presented in
\cite{Cai:2010zh}. New ideas concerning RG improvements have been proposed in
\cite{Becker:2012js,Becker:2012jx,Emoto:2005te,Ward:2006vw,Falls:2010he,Basu:2010nf,Casadio:2010fw}.

There are also attempts to investigate the nature of the microscopic structure of the asymptotically
safe quantum spacetime \cite{Lauscher:2005qz,Reuter:2011ah,Rechenberger:2012pm,Calcagni:2013vsa}.
These works reveal that quantum corrections at high energies (near the nontrivial UV fixed points)
modify drastically the classical picture. In this direction, but from another perspective, is the
study of the possibility to have singularity free gravitational collapse \cite{avoid}.

The truncated RG flow equations leave two running couplings (with respect to energy), the gravitational
constant $G(k)$ and the (positive) cosmological constant $\Lambda(k)$. Near the non-Gaussian UV
fixed point the field $G$ is known to approach zero, while on the other hand, the field
$\Lambda$ scales as the square of the momenta. The scope of our work is to investigate the
structure of spacetime at high energies and in particular the spherically symmetric configurations in
the asymptotic safety scenario in the vicinity of the center. This is why we have chosen to study
the modified Einstein vacuum equations where only $\Lambda(k)$ appears. In the same direction
is the observation \cite{Koch:2013owa,Koch:2013rwa} that the cosmological
constant seems to play a significant role in the understanding of the short distance behaviour.
In general, $\Lambda$ or $G$ energy dependent fields on the right hand side of Einstein
equations violate the Bianchi identities. In \cite{bianchi}, certain cases with consistent Bianchi
identities were presented and solutions were found, however, the case of our interest here
with $\Lambda$ a field unrelated to $G$ was not considered. In several attempts of
RG modified black hole solutions there is a violation of the Bianchi identities.

Here, we first find the appropriate covariant kinetic terms that support an arbitrary source field
$\Lambda(k)$ without any symmetry assumption and construct modified Einstein equations which
respect the Bianchi identities. Assuming the scaling of $\Lambda(k)$ proposed by the asymptotic safety
scenario at high energies, we write down and solve exactly the consistent quantum gravity improved
equations that determine the spherically symmetric solutions. A set of
different branches of solutions are obtained. The study of their curvature invariants and their
geodesic equations reveals a variety of smoothening behaviours in clear contrast to the singularities
accompanying classical General Relativity. Which of these solutions is the physically acceptable one
will be determined by the demand to match smoothly with the classical Schwarzschild solution at
larger length scales.

\section{Field equations with a varying cosmological constant}
\label{nkjk}

In this section we add a spacetime-dependent cosmological constant $\Lambda(x)$ in the
4-dimensional vacuum Einstein equations so that $G_{\mu\nu}=-\Lambda g_{\mu\nu}$. Although in the
scenario of asymptotic safety, $\Lambda(k)$ is supposed to be determined uniquely from the RG flow
equations, however, we want at this stage to construct a formulation where $\Lambda(x)$ can be kept
arbitrary (later we will specify the function $\Lambda(k)$). Obviously, the above equation is
inconsistent since it does not satisfy the Bianchi identities. We will add an energy-momentum tensor
$\vartheta_{\mu\nu}$ to support $\Lambda(x)$, so that the Bianchi identities are satisfied.
The new equations of motion become
\begin{equation}
G_{\mu\nu}=-\Lambda g_{\mu\nu}+\vartheta_{\mu\nu}\,,
\label{njg}
\end{equation}
or equivalently
\begin{equation}
R_{\mu\nu}=\Lambda g_{\mu\nu}+\vartheta_{\mu\nu}-\frac{1}{2}\vartheta\,g_{\mu\nu}\,,
\label{ajd}
\end{equation}
where $\vartheta=\vartheta^{\mu}_{\,\,\,\mu}$. The Bianchi identities take the form
\begin{equation}
\vartheta_{\mu\nu}^{\,\,\,\,\,\,\,;\mu}=\Lambda_{;\nu}\,,
\label{sws}
\end{equation}
where $;$ denotes covariant differentiation with respect to the Christoffel connection of $g_{\mu\nu}$.
Moreover, obviously $\vartheta_{\mu\nu}$ must vanish for $\Lambda$ constant since in this case the
Bianchi identities of (\ref{njg}) are trivially satisfied and there is no need to add anything.
Therefore, since the solution of the homogeneous equation $\vartheta_{\mu\nu}^{\,\,\,\,\,\,\,;\mu}=0$
must be $\vartheta_{\mu\nu}=0$, a partial solution of the inhomogeneous equation (\ref{sws}) will be
the general one.

It turns out to be very convenient to introduce the field $\psi(x)$ by
\begin{equation}
\Lambda=\bar{\Lambda}\,e^{\psi}\,,
\label{asj}
\end{equation}
where $\bar{\Lambda}$ is an arbitrary constant reference value.
We will construct such a tensor $\vartheta_{\mu\nu}$ from $\psi$ and its first and second derivatives.
We make the ansatz
\begin{equation}
\vartheta_{\mu\nu}=A(\psi)\psi_{;\mu}\psi_{;\nu}+B(\psi)g_{\mu\nu}\psi^{;\rho}
\psi_{;\rho}+C(\psi)\psi_{;\mu;\nu}+E(\psi)g_{\mu\nu}\Box\psi+F(\psi)g_{\mu\nu}\,,
\label{fsd}
\end{equation}
where $\Box\psi=\psi_{;\mu}^{\,\,\,\,\,;\mu}$. Then,
\begin{equation}
\vartheta=(A+4B)\psi^{;\mu}\psi_{;\mu}+(C+4E)\Box\psi+4F\,,
\label{jjs}
\end{equation}
\begin{equation}
R_{\mu\nu}=A\psi_{;\mu}\psi_{;\nu}-\Big(\frac{A}{2}+B\Big)g_{\mu\nu}\psi^{;\rho}\psi_{;\rho}
+C\psi_{;\mu;\nu}-\Big(\frac{C}{2}+E\Big)g_{\mu\nu}\Box\psi+(\bar{\Lambda}e^{\psi}-F)g_{\mu\nu}\,.
\label{hsh}
\end{equation}
Since $\Box(\psi_{;\nu})=(\Box\psi)_{;\nu}+R_{\nu\sigma}\psi^{;\sigma}$, it is from (\ref{fsd})
\begin{equation}
\vartheta_{\mu\nu}^{\,\,\,\,\,\,\,;\mu}=(A'\!+\!B')\psi^{;\mu}\psi_{;\mu}\psi_{;\nu}
+(A\!+\!E')\Box\psi\,\psi_{;\nu}+(C'\!+\!A\!+\!2B)\psi^{;\mu}\psi_{;\mu;\nu}
+(C\!+\!E)(\Box\psi)_{;\nu}+F'\psi_{;\nu}+CR_{\mu\nu}\psi^{;\mu}\,,
\label{wks}
\end{equation}
where a prime denotes differentiation with respect to $\psi$. Using (\ref{hsh}), equation (\ref{wks}) becomes
\begin{eqnarray}
\vartheta_{\mu\nu}^{\,\,\,\,\,\,\,;\mu}&\!\!\!\!=\!\!\!\!&\Big(A'\!+\!B'\!+\!\frac{1}{2}AC\!-\!BC\Big)
\psi^{;\mu}\psi_{;\mu}\psi_{;\nu}
+\Big(E'\!+\!A\!-\!CE\!-\!\frac{1}{2}C^{2}\Big)\Box\psi\,\psi_{;\nu}
+(C'\!+\!A\!+\!2B\!+\!C^{2})\psi^{;\mu}\psi_{;\mu;\nu}\nn\\
&&+(C\!+\!E)(\Box\psi)_{;\nu}+(F'\!-\!CF\!+\!C\bar{\Lambda}e^{\psi})\psi_{;\nu}\,.
\label{sjj}
\end{eqnarray}
The consistency condition (\ref{sws}) becomes
\begin{eqnarray}
&&\Big(A'\!+\!B'\!+\!\frac{1}{2}AC\!-\!BC\Big)
\psi^{;\mu}\psi_{;\mu}\psi_{;\nu}
+\Big(E'\!+\!A\!-\!CE\!-\!\frac{1}{2}C^{2}\Big)\Box\psi\,\psi_{;\nu}
+(C'\!+\!A\!+\!2B\!+\!C^{2})\psi^{;\mu}\psi_{;\mu;\nu}\nn\\
&&+(C\!+\!E)(\Box\psi)_{;\nu}+\big[F'\!-\!CF\!+\!(C-1)\bar{\Lambda}e^{\psi}\big]\psi_{;\nu}=0\,.
\label{ekk}
\end{eqnarray}
Since (\ref{ekk}) must be satisfied for any $\psi$, the various prefactors must vanish separately,
giving the following system of differential equations
\begin{eqnarray}
A'\!+\!B'\!+\!\frac{1}{2}AC\!-\!BC=0&\,\,,\,\,&E'\!+\!A\!-\!CE\!-\!\frac{1}{2}C^{2}=0
\,\,\,\,,\,\,\,\,C'\!+\!A\!+\!2B\!+\!C^{2}=0\nn\\
C\!+\!E=0&\,\,,\,\,&F'\!-\!CF\!+\!(C-1)\bar{\Lambda}e^{\psi}=0\,.
\label{jsd}
\end{eqnarray}
It can be seen that the first equation is redundant and the system gets the form
\begin{equation}
A+B=-\frac{3}{4}C^{2}\,\,\,\,,\,\,\,\,C'=\frac{1}{2}C^{2}+A\,\,\,\,,\,\,\,\,E=-C
\,\,\,\,,\,\,\,\,F'\!-\!CF\!+\!(C-1)\bar{\Lambda}e^{\psi}=0\,.
\label{jdn}
\end{equation}
The last equation in (\ref{jdn}) for $F'$ is a linear differential equation with solution
$F=e^{\int C}\big[\sigma+\bar{\Lambda}\int (1-C)e^{\psi-\int C}\big]$, where $\sigma$ is
integration constant. Since for $\psi$ constant, $\vartheta_{\mu\nu}$ must vanish, it arises from
(\ref{fsd}) that for $\psi$ constant the factor $F$ must vanish. From the form of the previous
solution for $F$, for this to happen for any constant $\psi$, it needs to be $C=1$ and furthermore
$\sigma=0$. Thus, although (\ref{jdn}) is a system of four equations for five unknowns, the
above requirement gives a unique solution. The system (\ref{jdn}) has therefore the solution
\begin{equation}
A=-\frac{1}{2}\,\,\,\,,\,\,\,\,B=-\frac{1}{4}\,\,\,\,,\,\,\,C=1\,\,\,\,,\,\,\,\,
E=-1\,\,\,\,,\,\,\,\,F=0\,.
\label{hdk}
\end{equation}
Finally $\vartheta_{\mu\nu}$ takes the form
\begin{equation}
\vartheta_{\mu\nu}=-\frac{1}{2}\psi_{;\mu}\psi_{;\nu}-\frac{1}{4}g_{\mu\nu}\psi^{;\rho}
\psi_{;\rho}+\psi_{;\mu;\nu}-g_{\mu\nu}\Box\psi\,.
\label{djm}
\end{equation}
Equation (\ref{djm}) is the general solution of (\ref{sws}) and could also be expressed in terms of
$\Lambda$. The self-consistent modified Einstein equations (\ref{njg}) become
\begin{equation}
G_{\mu\nu}=-\bar{\Lambda}\,e^{\psi} g_{\mu\nu}
-\frac{1}{2}\psi_{;\mu}\psi_{;\nu}-\frac{1}{4}g_{\mu\nu}\psi^{;\rho}
\psi_{;\rho}+\psi_{;\mu;\nu}-g_{\mu\nu}\Box\psi\,,
\label{wuj}
\end{equation}
containing an external field $\psi(x)$ which is given by a favored theory. Note that $\psi$
does not have its own equation of motion.

If an extra matter energy-momentum tensor $T_{\mu\nu}$ is included (which will not be the case in the
present work) the field equations become
\begin{equation}
G_{\mu\nu}=-\bar{\Lambda}\,e^{\psi} g_{\mu\nu}
-\frac{1}{2}\psi_{;\mu}\psi_{;\nu}-\frac{1}{4}g_{\mu\nu}\psi^{;\rho}
\psi_{;\rho}+\psi_{;\mu;\nu}-g_{\mu\nu}\Box\psi+8\pi G T_{\mu\nu}\,,
\label{skl}
\end{equation}
where the gravitational constant $G$ is in general also spacetime dependent. Since the various
$\psi$-kinetic terms along with the $\Lambda$ field in (\ref{skl}) are by construction covariantly
conserved, the Bianchi identities provide that the energy-momentum is not separately conserved but
only in combination with $G$, thus
\begin{equation}
(GT_{\mu\nu})^{;\nu}=0\,.
\label{ewi}
\end{equation}
If one insisted on the exact energy-momentum conservation, then extra covariant terms constructed
out of $G$ would be necessary to be added in (\ref{skl}). However, we do not think that it is
pertinent to include such an extra assumption. Therefore, since the RG flow equations aim to provide
energy dependent $\Lambda(k), G(k)$, the system of equations (\ref{skl}), (\ref{ewi}) is a
consistent system of equations valid at any energy scale in the presence of matter.
The consistency closes with the connection of the momentum scale $k$ with some length scale and the
only limitation is the restricted knowledge possessed about the form of the functions $\Lambda(k),
G(k)$ as a result of the RG flow equations.

There is another approach where the quantum corrections of $G$ and $\Lambda$ enter not through the
improved field equations (\ref{skl}), but through an improved action whose variation gives different
field equations. However, in this approach the full $\Lambda$ and $G$-dependent terms which make
the equations consistent have not been found in general. As for equations (\ref{skl}) it is not
evident if they arise from an action or not.

\section{Spherical symmetry}
\label{sphesy}

We are looking for static spherically symmetric solutions. The general metric is of the form
\begin{equation}
ds^{2}=-A(r)dt^{2}+B(r)dr^{2}+r^{2} (d \theta^{2}+\sin^{2}{\!\theta}\,d\phi^{2})\,.
\label{jkw}
\end{equation}
The external field $\Lambda(x)$ carries the same symmetries, so it is $\Lambda=\Lambda(r)$. For the
metric (\ref{jkw}) the non-vanishing components of the Einstein tensor $G_{\mu}^{\nu}$ are
\begin{eqnarray}
G^{t}_{t}&=&\frac{1}{r^{2}B}-\frac{1}{r^{2}}-\frac{B'}{rB^{2}}\\
G^{r}_{r}&=&\frac{1}{r^{2}B}-\frac{1}{r^{2}}+\frac{A'}{rAB}\\
G^{\theta}_{\theta}=G^{\phi}_{\phi}&=&\frac{A'}{2rAB}-\frac{B'}{2rB^{2}}-\frac{A'^{2}}{4A^{2}B}-
\frac{A'B'}{4AB^{2}}+\frac{A''}{2AB}\,.
\label{skk}
\end{eqnarray}
A prime denotes here and in the following differentiation with respect to $r$.
It is $\psi^{;\rho}\psi_{;\rho}=\frac{\psi'^{2}}{B}$, $\Box\psi=\frac{2\psi'}{rB}+\frac{A'\psi'}{2AB}
-\frac{B'\psi'}{2B^{2}}+\frac{\psi''}{B}$ and the non-vanishing components of $\psi_{;\mu;\nu}$ are
$\psi_{;t;t}=-\frac{A'\psi'}{2B}$, $\psi_{;r;r}=\psi''-\frac{B'\psi'}{2B}$, $\psi_{;\theta;\theta}=
\frac{r\psi'}{B}$, $\psi_{;\phi;\phi}=\frac{r\sin^{2}{\!\theta}\,\psi'}{B}$. Therefore, the three independent
components $tt,rr,\theta\theta$ of (\ref{wuj}) are
\begin{eqnarray}
\frac{1-B}{r^{2}}-\frac{B'}{rB}&=&-\bar{\Lambda}e^{\psi}B
-\frac{\psi'^{2}}{4}-\frac{2\psi'}{r}+\frac{B'\psi'}{2B}-\psi''
\label{wkd}\\
\frac{1-B}{r^{2}}+\frac{A'}{rA}&=&-\bar{\Lambda}e^{\psi}B
-\frac{3\psi'^{2}}{4}-\frac{2\psi'}{r}-\frac{A'\psi'}{2A}
\label{kee}\\
\frac{A'}{2rA}-\frac{B'}{2rB}-\frac{A'^{2}}{4A^{2}}-
\frac{A'B'}{4AB}+\frac{A''}{2A}&=&-\bar{\Lambda}e^{\psi}B-\frac{\psi'^{2}}{4}
-\frac{\psi'}{r}-\frac{A'\psi'}{2A}+\frac{B'\psi'}{2B}-\psi''\,.
\label{sdf}
\end{eqnarray}
The system of equations (\ref{wkd})-(\ref{sdf}) is satisfied by construction for any $\psi(r)$.
It can be seen that the equation (\ref{sdf}) is redundant, since it arises from (\ref{wkd}),
(\ref{kee}) and this reflects the $r-$reparametrization invariance of the system (\ref{wuj}).
Indeed, differentiating (\ref{kee}) with respect to $r$, we obtain together with
(\ref{wkd}) an algebraic system for $A'',\psi''$. We solve this system for $A'',\psi''$ and
we also solve (\ref{kee}) for $A'$. Plugging these $A'',\psi'',A'$ in (\ref{sdf}) we find
an identification.

At the UV the RG flow equations provide $\Lambda=\lambda_{\ast}k^{2}$, where $\lambda_{\ast}$ is
a dimensionless constant. Note that this scaling is also
consistent with dimensional analysis without the introduction of a new energy scale.
It is natural to set a relation between the energy and length scale of the form \cite{Bonanno:2001xi}
\begin{equation}
k=\frac{\xi}{D}\,,
\label{ppp}
\end{equation}
where $\xi$ is a dimensionless parameter and $D=D(r)>0$ is
the proper length of a radial curve defined by $dt=d\theta=d\phi=0$ connecting a point with coordinate
$r$ to the minimum radius of definition $r=r_{\text{min}}$ (possibly the origin $r_{\text{min}}=0$).
Thus,
\begin{equation}
D(r)=\int_{r_{\text{min}}}^{r}\sqrt{|B(\rho)|}\,d\rho
\label{qsl}
\end{equation}
and $D'(r)=\sqrt{|B(r)|}$.
It turns out that $\psi'=-2\frac{D'}{D}$, $\frac{B'}{B}=2\frac{D''}{D'}$. Equation (\ref{wkd}) is
written as
\begin{equation}
2\frac{D''}{D'}+\frac{1}{r}\Big(\epsilon-\alpha\frac{r^{2}}{D^{2}}\Big)D'^{2}
+4\frac{D'}{D}-\frac{1}{r}=0\,,
\label{jwd}
\end{equation}
where $\alpha=3+\epsilon \lambda_{\ast}\xi^{2}$ and $\epsilon$ is the sign of $B(r)$.
Normally it is $\epsilon=+1$, but since the signature of the metric is not known close to the
centres which are the regions of high energy, we leave $\epsilon=-1$ as a possibility (for example,
the interior Schwarzschild black hole in the same spherical coordinates has $\epsilon=-1$). As for
$A$ however, we will require in the following that $A$ and $B$ have the same signs, so there is
always a single time dimension. Of course, $r_{\text{min}}$ does not enter (\ref{jwd}),
or to say it different, an arbitrary constant could have been added on the right hand side
of (\ref{qsl}) without changing (\ref{jwd}). This means that in the end, the normalization
in order to find $D$ will come by demanding $D=0$ for $r=r_{\text{min}}$.

Equation (\ref{jwd}) has the symmetry $D\rightarrow \lambda D$,
$r\rightarrow \lambda r$, so under this transformation, (\ref{jwd}) remains invariant. Setting
\begin{equation}
u=\frac{D}{r}\,,
\label{ref}
\end{equation}
equation (\ref{jwd}) becomes
\begin{equation}
2\frac{ru''\!+\!2u'}{ru'\!+\!u}+\frac{1}{r}\Big(\epsilon-\frac{\alpha}{u^{2}}\Big)(ru'\!+\!u)^{2}
+4\frac{ru'\!+\!u}{ru}-\frac{1}{r}=0\,.
\label{erg}
\end{equation}
Under the redefinition
\begin{equation}
r=e^{\tau}\,,
\label{gef}
\end{equation}
equation (\ref{erg}) becomes
\begin{equation}
2\frac{\ddot{u}+\dot{u}}{\dot{u}+u}+\Big(\epsilon-\frac{\alpha}{u^{2}}\Big)
(\dot{u}+u)^{2}+4\frac{\dot{u}+u}{u}-1=0\,,
\label{lef}
\end{equation}
where a dot denotes differentiation with respect to $\tau$. Defining
\begin{equation}
v=\dot{u}\,,
\label{fed}
\end{equation}
equation (\ref{lef}) becomes
\begin{equation}
\frac{2v}{v+u}\frac{d}{du}(v+u)+\Big(\epsilon-\frac{\alpha}{u^{2}}\Big)(v+u)^{2}+4\frac{v+u}{u}-1=0\,.
\label{kef}
\end{equation}
Setting
\begin{equation}
Y=v+u\,,
\label{dwe}
\end{equation}
equation (\ref{kef}) reduces to the first order equation
\begin{equation}
2(Y\!-\!u)\frac{dY}{du}+\Big(\epsilon-\frac{\alpha}{u^{2}}\Big)Y^{3}+\frac{4}{u}Y^{2}-Y=0\,.
\label{weg}
\end{equation}
Under the replacement
\begin{equation}
y=\frac{1}{Y}\,,
\label{dvg}
\end{equation}
equation (\ref{weg}) becomes
\begin{equation}
\Big(y-\frac{1}{u}\Big)\frac{dy}{du}=\frac{1}{2u}y^{2}-\frac{2}{u^{2}}y+\frac{1}{2u}\Big(\frac{\alpha}{u^{2}}-
\epsilon\Big)\,.
\label{ddh}
\end{equation}
Equation (\ref{ddh}) can be converted to an Abel equation and such an equation is not in general solvable.
However, very luckily,
here the coefficients of (\ref{ddh}) are of a special form and equation (\ref{ddh}) can be converted to a
separable equation.
Namely, setting
\begin{equation}
z=\frac{1}{\sqrt{u}}\Big(y-\frac{1}{u}\Big)\,,
\label{weg}
\end{equation}
equation (\ref{ddh}) becomes
\begin{equation}
z\frac{dz}{du}=\frac{\alpha-3}{2u^{4}}-\frac{\epsilon}{2u^{2}}\,,
\label{fwg}
\end{equation}
with general solution
\begin{equation}
z=\pm \sqrt{\frac{\beta}{u^{3}}+\frac{\epsilon}{u}+c}\,\,\,\,\,\,,\,\,\,\,\,\,\beta=1-\frac{\alpha}{3}\,,
\label{seg}
\end{equation}
where $c$ is integration constant. Then,
\begin{equation}
y=\frac{1\pm\sqrt{cu^{3}\!+\!\epsilon u^{2}\!+\!\beta}}{u}\,.
\label{jwe}
\end{equation}
It is obvious that it has to be $cu^{3}\!+\!\epsilon u^{2}\!+\!\beta\geq 0$.

From the above definitions of the various variables, the metric component $B(u)=B(r(u))$ is found to be
\begin{equation}
B(u)=\frac{\epsilon u^{2}}{\big(1\!\pm\!\sqrt{cu^{3}\!+\!\epsilon u^{2}\!+\!\beta}\big)^{2}}\,.
\label{hre}
\end{equation}
In order to find the metric component $A$, it is convenient to introduce the variable $\chi$ instead
of $A$ defined by
\begin{equation}
A=\frac{1}{B}e^{2\chi}\,.
\label{jef}
\end{equation}
Plugging this expression in equation (\ref{kee}) and subtracting from (\ref{wkd}) we get
\begin{equation}
\Big(\psi'\!+\!\frac{2}{r}\Big)\chi'=\psi''\!-\!\frac{1}{2}\psi'^{2}\,.
\label{fsw}
\end{equation}
Since $u'=\frac{1}{r}(\frac{1}{y}-u)$, $\psi'=-\frac{2}{ruy}$, it is
$\psi''=\frac{2}{r^{2}uy^{2}}[\frac{1}{u}+(\frac{1}{y}-u)\frac{dy}{du}]$. Then, equation (\ref{fsw})
due to (\ref{ddh}) becomes
\begin{equation}
2(1\!-\!uy)^{2}\,\frac{d\chi}{du}=y+\Big(\frac{\alpha}{u^{2}}\!-\!\epsilon\Big)\frac{1}{y}-\frac{4}{u}\,.
\label{sdg}
\end{equation}
Replacing the solution (\ref{jwe}) in (\ref{sdg}) we obtain the equation
\begin{equation}
2\Big(cu^{3}\!+\!\epsilon u^{2}\!+\!1\!-\!\frac{\alpha}{3}\Big)\,\frac{d\chi}{du}
=-\frac{3cu^{2}\!+\!2\epsilon u}{cu^{3}\!+\!\epsilon u^{2}\!-\!\frac{\alpha}{3}}
\pm \frac{\big(cu^{3}\!+\!\frac{2\alpha}{3}\big)
\sqrt{cu^{3}\!+\!\epsilon u^{2}\!+\!1\!-\!\frac{\alpha}{3}}}
{u(cu^{3}\!+\!\epsilon u^{2}\!-\!\frac{\alpha}{3})}\,.
\label{ejv}
\end{equation}
Integrating this equation we find
\begin{equation}
2\chi(u)=\ln{\vartheta}+\ln{\frac{cu^{3}\!+\!\epsilon u^{2}\!+\!1\!-\!\frac{\alpha}{3}}
{\big|cu^{3}\!+\!\epsilon u^{2}\!-\!\frac{\alpha}{3}\big|}}\pm \texttt{J}(u)\,,
\label{weg}
\end{equation}
where $\vartheta>0$ is integration constant (not to be confused with $\vartheta^{\mu}_{\,\,\,\mu}$
of section \ref{nkjk}) and
\begin{equation}
\texttt{J}(u)=\int\!\frac{cu^{3}\!+\!\frac{2\alpha}{3}}{u(cu^{3}\!+\!\epsilon u^{2}\!-\!\frac{\alpha}{3})
\sqrt{cu^{3}\!+\!\epsilon u^{2}\!+\!1\!-\!\frac{\alpha}{3}}}du\,.
\label{kjc}
\end{equation}
Then, $A$ is found as
\begin{equation}
A(u)=\frac{\epsilon\vartheta}{u^{2}}\,
\Bigg(\!\!1\!\pm\!\sqrt{cu^{3}\!+\!\epsilon u^{2}\!+\!1\!-\!\frac{\alpha}{3}}\Bigg)^{\!\!2}\,\,
\frac{cu^{3}\!+\!\epsilon u^{2}\!+\!1\!-\!\frac{\alpha}{3}}
{\big|cu^{3}\!+\!\epsilon u^{2}\!-\!\frac{\alpha}{3}\big|}\,e^{\pm \texttt{J}(u)}\,,
\label{egh}
\end{equation}
where $A(u)=A(r(u))$.
Finally, the relation between $u$ and $r$ is found from equation $ru'=\frac{1-uy}{y}$ to be
\begin{equation}
r=r_{0}e^{\,\int \!\frac{y}{1-uy}du}\,,
\label{kwg}
\end{equation}
where $r_{0}>0$ is integration constant. Using the solution (\ref{jwe}) we find
\begin{equation}
r(u)=\frac{r_{0}}{u}\,e^{\mp \mathcal{I}(u)}\,,
\label{ekg}
\end{equation}
where
\begin{equation}
\mathcal{I}(u)=\int\!\frac{du}{u\sqrt{cu^{3}\!+\!\epsilon u^{2}\!+\!1\!-\!\frac{\alpha}{3}}}\,.
\label{efl}
\end{equation}

To summarize, the metric in the non-conventional gauge determined by $u$ is given by
\begin{equation}
ds^{2}=-A(u)dt^{2}+r(u)^{2}\Big[\frac{\epsilon}{cu^{3}\!+\!\epsilon u^{2}\!+\!1\!-\!\frac{\alpha}{3}}
\,du^{2}+(d\theta^{2}+\sin^{2}{\!\theta}\,d\phi^{2})\Big]\,.
\label{abt}
\end{equation}

It is more convenient to write the metric (\ref{abt}) in the form
\begin{equation}
ds^{2}=\mathcal{R}(u)^{2}\Big[-\mathcal{A}(u)dt^{2}
+\frac{\epsilon r_{0}^{2}}{cu^{3}\!+\!\epsilon u^{2}\!+\!1\!-\!\frac{\alpha}{3}}
\,du^{2}+r_{0}^{2}(d\theta^{2}+\sin^{2}{\!\theta}\,d\phi^{2})\Big]\,,
\label{wef}
\end{equation}
where
\begin{equation}
\mathcal{R}(u)=\frac{1}{u}\,e^{\mp \mathcal{I}(u)}
\label{elr}
\end{equation}
\begin{equation}
\mathcal{A}(u)=\epsilon\vartheta
\Bigg(\!\!1\!\pm\!\sqrt{cu^{3}\!+\!\epsilon u^{2}\!+\!1\!-\!\frac{\alpha}{3}}\Bigg)^{\!\!2}\,\,
\frac{cu^{3}\!+\!\epsilon u^{2}\!+\!1\!-\!\frac{\alpha}{3}}
{\big|cu^{3}\!+\!\epsilon u^{2}\!-\!\frac{\alpha}{3}\big|}\,e^{\pm \mathcal{J}(u)}
\label{wqk}
\end{equation}
\begin{equation}
\mathcal{J}(u)=\int\!\frac{3cu^{2}\!+\!2\epsilon u}{(cu^{3}\!+\!\epsilon u^{2}\!-\!\frac{\alpha}{3})
\sqrt{cu^{3}\!+\!\epsilon u^{2}\!+\!1\!-\!\frac{\alpha}{3}}}du\,.
\label{qwl}
\end{equation}
The relation of $r(u)$ and $\mathcal{R}(u)$ is $r(u)=r_{0}\mathcal{R}(u)$, while the
relation of $A(u)$ and $\mathcal{A}(u)$ is $A(u)=\frac{1}{u^{2}}\mathcal{A}(u)e^{\mp 2\mathcal{I}(u)}$.

It is useful to define the variable
\begin{equation}
\textmd{v}=u+\frac{\epsilon}{3c}\,.
\label{ewk}
\end{equation}
Since $u>0$, it should be $\textmd{v}>\frac{\epsilon}{3c}$.
The metric (\ref{wef}) is expressed in terms of $\textmd{v}$ as
\begin{equation}
ds^{2}=\textsf{R}(\textmd{v})^{2}\Big[-\textsf{A}(\textmd{v})dt^{2}
+\frac{\epsilon r_{0}^{2}}{c(\textmd{v}^{3}\!+\!p\textmd{v}\!+\!q)}
\,d\textmd{v}^{2}+r_{0}^{2}(d\theta^{2}+\sin^{2}{\!\theta}\,d\phi^{2})\Big]\,,
\label{ewc}
\end{equation}
where
\begin{equation}
\textsf{R}(\textmd{v})=\frac{1}{\textmd{v}\!-\!\frac{\epsilon}{3c}}\,e^{\mp I(\textmd{v})}
\label{afa}
\end{equation}
\begin{equation}
\textsf{A}(\textmd{v})=\epsilon\,\vartheta\,\,\text{sgn}(c)
\Big[1\!\pm\!\sqrt{c(\textmd{v}^{3}\!+\!p\textmd{v}\!+\!q)}\Big]^{2}\,\,
\frac{\textmd{v}^{3}\!+\!p\textmd{v}\!+\!q}
{\big|\textmd{v}^{3}\!+\!p\textmd{v}\!+\!q\!-\!\frac{1}{c}\big|}\,e^{\pm J(\textmd{v})}\,.
\label{vsa}
\end{equation}
The integrals $I(\textmd{v}),J(\textmd{v})$ are
\begin{equation}
I(\textmd{v})
=\int\!\frac{d\textmd{v}}{\big(\textmd{v}\!-\!\frac{\epsilon}{3c}\big)\sqrt{P(\textmd{v})}}
\label{jcd}
\end{equation}
\begin{equation}
J(\textmd{v})=\int\!\frac{3\textmd{v}^{2}\!+\!p}
{\big(\textmd{v}^{3}\!+\!p\textmd{v}\!+\!q\!-\!\frac{1}{c}
\big)\sqrt{P(\textmd{v})}}d\textmd{v}\,,
\label{ekl}
\end{equation}
where
\begin{equation}
P(\textmd{v})=c(\textmd{v}^{3}\!+\!p\textmd{v}\!+\!q)
\label{iwj}
\end{equation}
\begin{equation}
p=-\frac{1}{3c^{2}}\,\,\,\,\,\,\,\,\,,\,\,\,\,\,\,\,\,\,
q=\frac{1}{c}\Big(\frac{2\epsilon}{27c^{2}}\!+\!1\!-\!\frac{\alpha}{3}\Big)\,.
\label{owe}
\end{equation}
Now, it is $r(\textmd{v})=r_{0}\textsf{R}(\textmd{v})$ and what remains in the next is the calculation
of the integrals $I(\textmd{v}),J(\textmd{v})$ encountered in
$\textsf{R}(\textmd{v}),\textsf{A}(\textmd{v})$. From (\ref{ref}), (\ref{ekg}) and since
$\mathcal{I}(u)=I(\textmd{v})$ we get
\begin{equation}
D=r_{0} \big[e^{\mp I(\textmd{v})}-e^{\mp I(\hat{\textmd{v}})}\big]\,,
\label{jem}
\end{equation}
where $\hat{\textmd{v}}$ corresponds to $r_{\text{min}}$.

Note that $\textsf{A}(\textmd{v})$ from (\ref{vsa}) is written as
\begin{equation}
\textsf{A}(\textmd{v})=\epsilon\vartheta\frac{\big(\sqrt{P(\textmd{v})}\!\pm\!1\big)^{2}}
{|P(\textmd{v})^{-1}\!-\!1|}e^{\pm J(\textmd{v})}\,,
\label{slk}
\end{equation}
while from (\ref{ekl}) it arises that
\begin{equation}
\frac{dJ}{d\textmd{v}}=\frac{1}{\sqrt{P}\,(P\!-\!1)}\frac{dP}{d\textmd{v}}\,.
\label{oen}
\end{equation}
Differentiating (\ref{slk}) with respect to $\textmd{v}$ and using (\ref{oen})
we obtain the simple differential equation
\begin{equation}
\frac{1}{\textsf{A}}\frac{d\textsf{A}}{d\textmd{v}}=\frac{1}{P}\frac{dP}{d\textmd{v}}.
\label{jej}
\end{equation}
The solution of (\ref{jej}) is
\begin{equation}
\textsf{A}(\textmd{v})=\epsilon\zeta P(\textmd{v})
=\epsilon\zeta c(\textmd{v}^{3}\!+\!p\textmd{v}\!+\!q)\,,
\label{ldd}
\end{equation}
where $\zeta>0$ is integration constant. Therefore, although the integral $J(\textmd{v})$ in
(\ref{ekl}) looks rather complicated and can be converted to typical elliptic integrals, however,
the specific combination
$3\textmd{v}^{2}\!+\!p=\frac{1}{c}\frac{dP}{d\textmd{v}}$ in the numerator of (\ref{ekl})
implies that $e^{\pm J(\textmd{v})}$ is expressed as a simple algebraic function of $\textmd{v}$.
Finally, the spacetime metric (\ref{ewc}) takes the form
\begin{equation}
ds^{2}=\textsf{R}(\textmd{v})^{2}\Big[-\epsilon\zeta c (\textmd{v}^{3}\!+\!p\textmd{v}\!+\!q) dt^{2}
+\frac{\epsilon \,r_{0}^{2}}{c(\textmd{v}^{3}\!+\!p\textmd{v}\!+\!q)}
\,d\textmd{v}^{2}+r_{0}^{2}(d\theta^{2}+\sin^{2}{\!\theta}\,d\phi^{2})\Big]\,.
\label{wel}
\end{equation}
So, the problem has been reduced to the calculation of the integral $I(\textmd{v})$
determining $\textsf{R}(\textmd{v})$.
The solution (\ref{wel}) contains three integration constants $r_{0},c,\zeta$ but $\zeta$
can be absorbed in a rescaling of the time $t$.
Using (\ref{afa}), (\ref{jcd}), the Ricci scalar $\text{Ric}$
of the metric (\ref{wel}) can be found to be
\begin{eqnarray}
\text{Ric}&\!\!=\!\!&-\frac{6\epsilon c}{r_{0}^{2}}e^{\pm 2 I(\textmd{v})}\Big[
\pm\frac{1}{2\sqrt{P}}\Big(3\textmd{v}^{3}\!+\!\frac{\epsilon}{c}\textmd{v}^{2}\!+\!5p\textmd{v}
\!+\!6q\!+\!\frac{\epsilon p}{3c}\Big)\!+\!2q\!+\!\frac{4\epsilon p}{9c}\!+\!\frac{1}{c}\Big]
\label{wfj}\\
&\!\!=\!\!&-\frac{6\epsilon c}{r_{0}^{2}}e^{\pm 2 I(\textmd{v})}\Big[
\pm\frac{1}{2\sqrt{P}}\Big(\frac{\epsilon}{c}\textmd{v}^{2}\!+\!2p\textmd{v}
\!+\!3q\!+\!\frac{\epsilon p}{3c}\Big)\!+\!2q\!+\!\frac{4\epsilon p}{9c}\!+\!\frac{1}{c}
\!\pm\!\frac{3}{2c}\sqrt{P}\Big]\,.
\label{wfe}
\end{eqnarray}

\section{Solutions}
\label{Solns}

\subsection{General solutions}
\label{gen sols}

Up to this point, we have been quite general. Now, we will restrict our interest to the case where the
polynomial $P(\textmd{v})$ has three real roots $\rho_{0},\rho_{1},\rho_{2}$, i.e.
$\rho_{\ell}^{3}\!+\!p\rho_{\ell}\!+\!q=0$, $\ell=0,1,2$. Then $P(\textmd{v})$ is decomposed into
three simple factors without the need to introduce complex roots. The three real roots exist
when $4p^{3}+27q^{2}\leq 0$ (i.e. $|qc^{3}|\leq 2/27$), or equivalently
$|\frac{2\epsilon}{c^{2}}+9(3-\alpha)|\leq\frac{2}{c^{2}}$, which means
$0\leq \epsilon(\alpha-3)\leq \frac{4}{9c^{2}}$, and therefore
$0<\lambda_{\star}\xi^{2}\leq\frac{4}{9c^{2}}$. The roots $\rho_{\ell}$ are given by
\begin{equation}
\rho_{\ell}=\frac{2}{3|c|}
\cos{\Big[\frac{1}{3}\arccos{\Big(\!\!-\frac{27}{2}|c|^{3}q\Big)}+\frac{2\pi \ell}{3}\Big]}\,.
\label{wej}
\end{equation}
It can be seen that
$-\frac{2}{3|c|}\leq\rho_{1}\leq - \frac{1}{2}\frac{2}{3|c|}\leq\rho_{2}
\leq\frac{1}{2}\frac{2}{3|c|}\leq\rho_{0}\leq \frac{2}{3|c|}$. The variable $\textmd{v}$ is
defined in some interval (finite or infinite) such that $P(\textmd{v})\geq 0$ (formally the
equality should not be included as $P(\textmd{v})$ is denominator in (\ref{wel}), but it more
convenient to leave this notation in the following analysis).
Depending on the domain of definition of $\textmd{v}$, the integral $I(\textmd{v})$ may possess
an integrable pole $\textmd{v}=\rho_{\ell}$ and/or the non-integrable
(logarithmic) pole $\textmd{v}=\frac{\epsilon}{3c}$.
There are only four cases for the domain where $\textmd{v}$ is
defined: (i) $c>0$, $\textmd{v}\geq\rho_{0}$,
(ii) $c>0$, $\epsilon=-1$, $-\frac{1}{3c}< \textmd{v}\leq\rho_{2}$,
(iii) $c<0$, $\epsilon=1$, $\rho_{2}\leq\textmd{v}\leq\rho_{0}$,
(iv) $c<0$, $\epsilon=-1$, $-\frac{1}{3c}< \textmd{v}\leq\rho_{0}$.
In the following analysis we make use of the identities $\rho_{0}\!+\!\rho_{1}\!+\!\rho_{2}=0$,
$\rho_{0}\rho_{1}\!+\!\rho_{0}\rho_{2}\!+\!\rho_{1}\rho_{2}=p$, $\rho_{0}\rho_{1}\rho_{2}=-q$.
\newline

$\bullet$ {\bf{Case (i): $c>0$, $\textmd{v}\geq\rho_{0}$}}.
\newline
\noindent
We use the transformation
\begin{equation}
x=\sqrt{\frac{\textmd{v}\!-\!\rho_{0}}{\textmd{v}\!-\!\rho_{2}}}\,,
\label{efw}
\end{equation}
where the lower limit of the integral $I(\textmd{v})$ in (\ref{jcd})
is $\textmd{v}=\rho_{0}$ and the upper $\textmd{v}$.
We find (correcting the formula 7, p. 262 of \cite{GradR} and the formula 5, p. 58 of \cite{Prud})
\begin{equation}
I(\textmd{v})\!=\!\frac{2}{(\rho_{2}\!-\!\frac{\epsilon}{3c})
\sqrt{\!c(\rho_{0}\!-\!\rho_{1})}}
\Big[\hat{F}\Big(\!
\sqrt{\frac{\textmd{v}\!-\!\rho_{0}}{\textmd{v}\!-\!\rho_{2}}},
\frac{\rho_{2}\!-\!\rho_{1}}{\rho_{0}\!-\!\rho_{1}}\Big)\!-\!
\frac{\rho_{0}\!-\!\rho_{2}}{\rho_{0}\!-\!\frac{\epsilon}{3c}}
\,\hat{\Pi}\Big(\!\sqrt{\frac{\textmd{v}\!-\!\rho_{0}}{\textmd{v}\!-\!\rho_{2}}},
\frac{\rho_{2}\!-\!\frac{\epsilon}{3c}}{\rho_{0}\!-\!\frac{\epsilon}{3c}},
\frac{\rho_{2}\!-\!\rho_{1}}{\rho_{0}\!-\!\rho_{1}}\Big)\Big]\,,
\label{wej}
\end{equation}
where
\begin{equation}
\hat{F}(z,m)=\int_{0}^{z}
\!\frac{dx}{\sqrt{(1\!-\!x^{2})(1\!-\!mx^{2})}}
\label{wej}
\end{equation}
is the incomplete elliptic integral of the first kind, and
\begin{equation}
\hat{\Pi}(z,n,m)=\int_{0}^{z}
\!\frac{dx}{(1\!-\!nx^{2})\sqrt{(1\!-\!x^{2})(1\!-\!mx^{2})}}
\label{eln}
\end{equation}
is the incomplete elliptic integral of the third kind
\footnote{In the literature, $\hat{F}(z,m)$
is denoted by $F(\arcsin{z},\sqrt{m})$ in \cite{GradR}, \cite{Prud} and by $F(\arcsin{z}|m)$ in
\cite{AbraS} and Wolfram's Mathematica, while
$\hat{\Pi}(z,n,m)$ is denoted by $\Pi(\arcsin{z},n,\sqrt{m})$ in \cite{GradR}, \cite{Prud},
by $\Pi(n;\arcsin{z}\backslash\alpha)$ (with $m=\sin^{2}\alpha$) in \cite{AbraS} and
by $\Pi(n;\arcsin{z}|m)$ in Wolfram's Mathematica.}.
The above expressions hold for $\rho_{0}\neq\frac{1}{3c}$, $\rho_{2}\neq \frac{\epsilon}{3c}$
which mean $c^{3}|q|\neq \frac{2}{27}$. In this case, the function $I(\textmd{v})$
ranges from zero value at the minimum value of $\textmd{v}=\rho_{0}$ up to a finite positive value
$I_{\infty}$ for $\textmd{v}\rightarrow +\infty$.
\newline
For the upper branch of the various $\pm$ signs, the radial function $\textsf{R}(\textmd{v})$
extends from the origin $r=0$ (for $\textmd{v}\rightarrow \hat{\textmd{v}}=+\infty$) to a finite
radius (for $\textmd{v}=\rho_{0}$). At this finite radius, the coordinate
$r_{\text{max}}$ is from (\ref{afa}) $r_{\text{max}}=\frac{r_{0}}{\rho_{0}-\frac{\epsilon}{3c}}$, while
the proper distance $D$ is from (\ref{jem}) $D_{\text{max}}=r_{0}(1-e^{- I_{\infty}})$,
normalizing this way the integration constant $r_{0}$.
The Ricci scalar, as it is seen from (\ref{wfj}), diverges at the origin signaling a curvature
singularity there. Additionally, the metric (\ref{wel}) close to the origin $r\simeq 0$ takes
the following form
\begin{equation}
ds_{0}^{2}\simeq -\epsilon \zeta c e^{-3I_{\!\infty}}\frac{r_{0}}{r}dt^{2}
+\frac{\epsilon}{c}e^{I_{\!\infty}}\frac{r}{r_{0}}dr^{2}+r^{2}
(d\theta^{2}+\sin^{2}{\!\theta}\,d\phi^{2})\,,
\label{wpf}
\end{equation}
from where it is also seen that the Ricci scalar diverges as $R\sim r^{-2}$, while
$R_{\mu\nu\kappa\lambda}R^{\mu\nu\kappa\lambda}\sim r^{-6}$ (for the Schwarzschild metric it is
$R=0$, $R_{\mu\nu\kappa\lambda}R^{\mu\nu\kappa\lambda}\sim r^{-6}$).
Close to the origin $r\simeq 0$ the proper distance is
$D\simeq \frac{1}{\sqrt{cr_{0}}}e^{\frac{1}{2}I_{\infty}}r^{3/2}$, the same scaling as
Schwarzschild. {\it{For $\epsilon=1$, although the solution is divergent at the origin,
we will show in the next section that this divergency is not felt by any infalling timelike
particle of arbitrary initial velocity. The reason is that the particle, due to a seemingly
repulsive force close to the origin, stops at some finite distance and then it moves outwards.}}
\newline
For the lower branch of the $\pm$ sign, the consistent solution (with $D>0$) ranges from a minimum
non-zero radius $r_{\text{min}}=\frac{r_{0}}{\rho_{0}-\frac{\epsilon}{3c}}$
for $\textmd{v}=\hat{\textmd{v}}=\rho_{0}$ to a maximum radius
$r_{\text{max}}=\frac{r_{0}}{\varrho_{0}-\frac{\epsilon}{3c}}e^{I(\varrho_{0})}$ for
$\textmd{v}=\varrho_{0}>\rho_{0}$.
It is $\varrho_{0}=
\frac{2}{3c}\cos{\{\frac{1}{3}\arccos{[\frac{27}{2}c^{3}(\frac{1}{c}-q)]}\}}$ for $\epsilon=1$,
$0<\lambda_{\ast}\xi^{2}\leq\frac{4}{9c^{2}}-3$ or for $\epsilon=-1$,
$3<\lambda_{\ast}\xi^{2}\leq \frac{4}{9c^{2}}$, while it is
$\varrho_{0}=
-\frac{2}{3c}\text{sgn}(cq\!-\!1)\cosh{\{\frac{1}{3}\arccos\!\text{h}{[\frac{27}{2}c^{3}(\frac{1}{c}-q)]}\}}$
for $\epsilon=1$,
$0\leq\frac{4}{9c^{2}}-3<\lambda_{\ast}\xi^{2}\leq\frac{4}{9c^{2}}$ or for $\epsilon=-1$,
$0<\lambda_{\ast}\xi^{2}<3\leq \frac{4}{9c^{2}}$.
The maximum proper distance is $D_{\text{max}}=r_{0}[e^{I(\varrho_{0})}-1]$. However, still here, the
Ricci scalar diverges at the minimum radius, as it is seen from (\ref{wfe}), and we do not analyze
this solution further.
\newline

$\bullet$ {\bf{Case (ii): $c>0$, $\epsilon=-1$, $-\frac{1}{3c}< \textmd{v}\leq\rho_{2}$}}.
\newline
\noindent
We use the transformation
\begin{equation}
x=\sqrt{\frac{\rho_{0}\!-\!\rho_{1}}{\rho_{2}\!-\!\rho_{1}}\,
\frac{\rho_{2}\!-\!\textmd{v}}{\rho_{0}\!-\!\textmd{v}}}\,,
\label{ewm}
\end{equation}
where the lower limit of the integral $I(\textmd{v})$ in (\ref{jcd})
is $\textmd{v}$ and the upper $\textmd{v}=\rho_{2}$.
We find (in agreement with formula 4, p. 262 of \cite{GradR} and formula 6, p. 65 of \cite{Prud})
\begin{equation}
I(\textmd{v})\!=\!\frac{2}{(\rho_{0}\!+\!\frac{1}{3c})
\sqrt{\!c(\rho_{0}\!-\!\rho_{1})}}
\Big[\hat{F}\Big(\!
\sqrt{\frac{\rho_{0}\!-\!\rho_{1}}{\rho_{2}\!-\!\rho_{1}}\,
\frac{\rho_{2}\!-\!\textmd{v}}{\rho_{0}\!-\!\textmd{v}}},
\frac{\rho_{2}\!-\!\rho_{1}}{\rho_{0}\!-\!\rho_{1}}\Big)-
\frac{\rho_{2}\!-\!\rho_{0}}{\rho_{2}\!+\!\frac{1}{3c}}
\,\hat{\Pi}\Big(\!\sqrt{\frac{\rho_{0}\!-\!\rho_{1}}{\rho_{2}\!-\!\rho_{1}}\,
\frac{\rho_{2}\!-\!\textmd{v}}{\rho_{0}\!-\!\textmd{v}}},
\frac{\rho_{2}\!-\!\rho_{1}}{\rho_{0}\!-\!\rho_{1}}\,
\frac{\rho_{0}\!+\!\frac{1}{3c}}{\rho_{2}\!+\!\frac{1}{3c}},
\frac{\rho_{2}\!-\!\rho_{1}}{\rho_{0}\!-\!\rho_{1}}\Big)\Big]\,.
\label{ekb}
\end{equation}
In this case, the function $I(\textmd{v})$ ranges from $+\infty$ at the minimum value
of $\textmd{v}=-\frac{1}{3c}$ up to the zero value for $\textmd{v}=\rho_{2}$.
\newline
For the upper branch of the various $\pm$ signs and $3<\lambda_{\ast}\xi^{2}\leq \frac{4}{9c^{2}}$,
the consistent solution
(with $D>0$) ranges from a minimum non-zero radius
\begin{equation}
r_{\text{min}}=\frac{r_{0}}{\varrho_{2}+\frac{1}{3c}}
e^{-I(\varrho_{2})}
\label{yyy}
\end{equation}
for $\textmd{v}=\hat{\textmd{v}}=\varrho_{2}=
\frac{2}{3c}\cos{\{\frac{1}{3}\arccos{[\frac{27}{2}c^{3}(\frac{1}{c}-q)]+\frac{4\pi}{3}}\}}<\rho_{2}$
to a finite maximum radius $r_{\text{max}}=\frac{r_{0}}{\rho_{2}+\frac{1}{3c}}$ for $\textmd{v}=\rho_{2}$.
The maximum proper distance $D$ is from (\ref{jem}) $D_{\text{max}}=r_{0}[1-e^{- I(\varrho_{2})}]$.
At the maximum distance, the Ricci scalar is seen from (\ref{wfe}) that diverges. However, this
is not important since the solution found is physically realistic close to the minimum radius
where the cosmological constant scales as $k^{2}$; at larger distances a deviation from the $k^{2}$
law occurs and the solution has to be modified there, while at even larger distances the solution
should reduce to the low energy Schwarzschild (or Schwarzschild-De Sitter) solution. {\it{The important
information carried by this solution is the fact that the present branch has at the minimum radius
finite the Ricci scalar as well as the other curvature invariants.}} At
radii close to the minimum, this solution can also be considered as physically realistic and
non-singular since the curvature invariants are finite. This is a success
of the quantum gravity scenario under discussion and the avoidance of singularities is a silent hope
of quantizing gravity.
\newline
Yet for the upper branch, there is another solution for $0<\lambda_{\ast}\xi^{2}<3\leq \frac{4}{9c^{2}}$
which extends from the origin $r=0$ for $\textmd{v}=\hat{\textmd{v}}=-\frac{1}{3c}$ to a finite
radius $r_{\text{max}}=\frac{r_{0}}{\rho_{2}+\frac{1}{3c}}$ for $\textmd{v}=\rho_{2}$. However, the
Ricci scalar diverges at the origin.
\newline
For the lower branch of the $\pm$ sign, the solution ranges from a minimum
non-zero radius for $\textmd{v}=\rho_{2}$ to a maximum radius for $\textmd{v}=-\frac{1}{3c}$.
The Ricci scalar diverges at the minimum radius.
\newline

$\bullet$ {\bf{Case (iii): $c<0$, $\epsilon=1$, $\rho_{2}\leq\textmd{v}\leq\rho_{0}$}}.
\newline
\noindent
We use the transformation
\begin{equation}
x=\sqrt{\frac{\rho_{0}\!-\!\rho_{1}}{\rho_{0}\!-\!\rho_{2}}\,
\frac{\textmd{v}\!-\!\rho_{2}}{\textmd{v}\!-\!\rho_{1}}}\,,
\label{ewm}
\end{equation}
where the lower limit of the integral $I(\textmd{v})$ in (\ref{jcd})
is $\textmd{v}=\rho_{2}$ and the upper $\textmd{v}$.
We find (in agreement with formula 5, p. 262 of \cite{GradR} and formula 6, p. 63 of \cite{Prud})
\begin{equation}
I(\textmd{v})\!=\!\frac{2}{(\rho_{1}\!-\!\frac{1}{3c})
\sqrt{\!c(\rho_{1}\!-\!\rho_{0})}}
\Big[\hat{F}\Big(\!
\sqrt{\frac{\rho_{0}\!-\!\rho_{1}}{\rho_{0}\!-\!\rho_{2}}\,
\frac{\textmd{v}\!-\!\rho_{2}}{\textmd{v}\!-\!\rho_{1}}},
\frac{\rho_{0}\!-\!\rho_{2}}{\rho_{0}\!-\!\rho_{1}}\Big)\!-\!
\frac{\rho_{2}\!-\!\rho_{1}}{\rho_{2}\!-\!\frac{1}{3c}}
\,\hat{\Pi}\Big(\!\sqrt{\frac{\rho_{0}\!-\!\rho_{1}}{\rho_{0}\!-\!\rho_{2}}\,
\frac{\textmd{v}\!-\!\rho_{2}}{\textmd{v}\!-\!\rho_{1}}},
\frac{\rho_{1}\!-\!\frac{1}{3c}}{\rho_{2}\!-\!\frac{1}{3c}}\,
\frac{\rho_{0}\!-\!\rho_{2}}{\rho_{0}\!-\!\rho_{1}},
\frac{\rho_{0}\!-\!\rho_{2}}{\rho_{0}\!-\!\rho_{1}}\Big)\Big]\,.
\label{ejf}
\end{equation}
The above expressions hold for $\rho_{1},\rho_{2}\neq\frac{1}{3c}$ which mean $c^{3}q
\neq \frac{2}{27}$.
In this case, the function $I(\textmd{v})$ ranges from zero value at the minimum value
of $\textmd{v}=\rho_{2}$ up to a finite positive value for $\textmd{v}=\rho_{0}$.
\newline
For the upper branch of the $\pm$ sign, the solution ranges from a minimum
non-zero radius for $\textmd{v}=\rho_{0}$ to a maximum radius for $\textmd{v}=\rho_{2}$.
The Ricci scalar diverges at the minimum radius.
\newline
For the lower branch of the various $\pm$ signs and
$0<\lambda_{\ast}\xi^{2}<\frac{4}{9c^{2}}-3$, the consistent solution
(with $D>0$) ranges from a minimum non-zero radius
\begin{equation}
r_{\text{min}}=\frac{r_{0}}{\varrho_{0}-\frac{1}{3c}}e^{I(\varrho_{0})}
\label{ooo}
\end{equation}
for
$\textmd{v}=\hat{\textmd{v}}=\varrho_{0}=
-\frac{2}{3c}\cos{\{\frac{1}{3}\arccos{[\frac{27}{2}c^{3}(q-\frac{1}{c})]}\}}<\rho_{0}$
to a finite maximum radius
$r_{\text{max}}=\frac{r_{0}}{\rho_{0}-\frac{1}{3c}}e^{I(\rho_{0})}$ for
$\textmd{v}=\rho_{0}$.
The maximum proper distance $D$ is from (\ref{jem})
$D_{\text{max}}=r_{0}[e^{I(\rho_{0})}-e^{I(\varrho_{0})}]$. {\it This branch is the second
branch we meet in the analysis of solutions where the curvature invariants at the minimum
radius are all finite and note that the signature of this metric is the ``correct'' one.}
\newline
Yet for the lower branch, there are another two solutions. The first one for
$0<\lambda_{\ast}\xi^{2}<\frac{4}{9c^{2}}-3$
extends from a minimum non-zero radius for $\textmd{v}=\rho_{2}$ to a finite
maximum radius for $\textmd{v}=\varrho_{2}=
-\frac{2}{3c}\cos{\{\frac{1}{3}\arccos{[\frac{27}{2}c^{3}(q-\frac{1}{c})]}+\frac{4\pi}{3}\}}
>\rho_{2}$ and has divergent Ricci scalar at the minimum radius.
The second one for
$0<\frac{4}{9c^{2}}-3<\lambda_{\ast}\xi^{2}<\frac{4}{9c^{2}}$
or $0<\lambda_{\ast}\xi^{2}<\frac{4}{9c^{2}}<3$
also extends from a minimum to a maximum radius for $\textmd{v}=\rho_{2}$ and
$\textmd{v}=\rho_{0}$ respectively and also has divergent Ricci scalar at the minimum radius.
\newline

$\bullet$ {\bf{Case (iv): $c<0$, $\epsilon=-1$, $-\frac{1}{3c}<\textmd{v}\leq\rho_{0}$}}.
\newline
\noindent
We use the transformation
\begin{equation}
x=\sqrt{\frac{\rho_{0}\!-\!\textmd{v}}{\rho_{0}\!-\!\rho_{2}}}\,,
\label{ewm}
\end{equation}
where the lower limit of the integral $I(\textmd{v})$ in (\ref{jcd})
is $\textmd{v}$ and the upper $\textmd{v}=\rho_{0}$.
We find (in agreement with formula 6, p. 262 of \cite{GradR} and formula 5, p. 61 of \cite{Prud})
\begin{equation}
I(\textmd{v})\!=\!\frac{2}{(\rho_{0}\!+\!\frac{1}{3c})
\sqrt{\!c(\rho_{1}\!-\!\rho_{0})}}
\,\hat{\Pi}\Big(\!\sqrt{\frac{\rho_{0}\!-\!\textmd{v}}{\rho_{0}\!-\!\rho_{2}}},
\frac{\rho_{0}\!-\!\rho_{2}}{\rho_{0}\!+\!\frac{1}{3c}},
\frac{\rho_{0}\!-\!\rho_{2}}{\rho_{0}\!-\!\rho_{1}}\Big)\,.
\label{ejf}
\end{equation}
This case is very similar to the case (ii). The function $I(\textmd{v})$ ranges from $+\infty$ at
the minimum value of $\textmd{v}=-\frac{1}{3c}$ up to the zero value for $\textmd{v}=\rho_{0}$.
\newline
For the upper branch of the various $\pm$ signs and $3<\lambda_{\ast}\xi^{2}\leq \frac{4}{9c^{2}}$,
the consistent solution
(with $D>0$) ranges from a minimum non-zero radius
\begin{equation}
r_{\text{min}}=\frac{r_{0}}{\varrho_{0}+\frac{1}{3c}}e^{-I(\varrho_{0})}
\label{eee}
\end{equation}
for $\textmd{v}=\hat{\textmd{v}}=\varrho_{0}=
-\frac{2}{3c}\cos{\{\frac{1}{3}\arccos{[\frac{27}{2}c^{3}(q-\frac{1}{c})]}\}}<\rho_{0}$
to a finite maximum radius $r_{\text{max}}=\frac{r_{0}}{\rho_{0}+\frac{1}{3c}}$ for $\textmd{v}=\rho_{0}$.
The maximum proper distance $D$ is from (\ref{jem}) $D_{\text{max}}=r_{0}[1-e^{- I(\varrho_{0})}]$.
At the maximum distance, the Ricci scalar is seen from (\ref{wfe}) that diverges.
{\it{However, at the minimum radius, the Ricci scalar and the other curvature invariants are finite
and this is the third case of a non-singular solution which does not extend to the origin.}}
\newline
Yet for the upper branch, there is another solution for $0<\lambda_{\ast}\xi^{2}<3\leq \frac{4}{9c^{2}}$
which extends from the origin $r=0$ for $\textmd{v}=\hat{\textmd{v}}=-\frac{1}{3c}$ to a finite
radius $r_{\text{max}}=\frac{r_{0}}{\rho_{0}+\frac{1}{3c}}$ for $\textmd{v}=\rho_{0}$. However, the
Ricci scalar diverges at the origin.
\newline
For the lower branch of the $\pm$ sign, the solution ranges from a minimum
non-zero radius for $\textmd{v}=\rho_{0}$ to a maximum radius for $\textmd{v}=-\frac{1}{3c}$.
The Ricci scalar diverges at the minimum radius.

\subsection{A special solution}

In this subsection we study a simpler special solution which arises when the integration constant
$c=0$ and is not included in the general solutions derived above. From equation (\ref{wef})
we take the solution
\begin{equation}
ds^{2}=\mathcal{R}(u)^{2}\Big[-\epsilon\vartheta
\Big(\epsilon u^{2}\!+\!1\!-\!\frac{\alpha}{3}\Big)dt^{2}
+\frac{\epsilon r_{0}^{2}}{\epsilon u^{2}\!+\!1\!-\!\frac{\alpha}{3}}
\,du^{2}+r_{0}^{2}(d\theta^{2}+\sin^{2}{\!\theta}\,d\phi^{2})\Big]\,,
\label{djv}
\end{equation}
where $\epsilon u^{2}+1-\frac{\alpha}{3}> 0$. The function $\mathcal{R}(u)=\frac{r(u)}{r_{0}}$
is given from (\ref{elr}) for $\epsilon=1$ by
\begin{equation}
\mathcal{R}(u)=\frac{1}{u}e^{\mp\frac{1}{\hat{\alpha}}\arccos{\frac
{\hat{\alpha}}{u}}}\,\,\,\,,\,\,\,\,u>\hat{\alpha}\,\,,\,\,
\hat{\alpha}=\sqrt{\frac{\alpha}{3}-1}\,,
\label{wei}
\end{equation}
while for $\epsilon=-1$ by
\begin{equation}
\mathcal{R}(u)=\frac{1}{u}\Big(\frac{\sqrt{\hat{\alpha}^{2}\!-\!u^{2}}
\!+\!\hat{\alpha}}{u}\Big)^{\pm\frac{1}{\hat{\alpha}}}
\,\,\,\,,\,\,\,\,u<\hat{\alpha}\,\,,\,\,\hat{\alpha}=\sqrt{1-\frac{\alpha}{3}}\,.
\label{eif}
\end{equation}
The Ricci scalar is found to be
\begin{equation}
\text{Ric}=-\frac{6}{r_{0}^{2}}e^{\pm 2\mathcal{I}(u)}
\Big(\frac{9\!-\!2\alpha}{3\epsilon}\pm\frac{2u^{2}\!-\!3\hat{\alpha}^{2}}
{\sqrt{|u^{2}\!-\!\hat{\alpha}^{2}|}}\Big)\,,
\label{aoc}
\end{equation}
where $\mathcal{I}(u)=\frac{1}{\hat{\alpha}}\arccos{\frac{\hat{\alpha}}{u}}$ for $\epsilon=1$,
while $\mathcal{I}(u)=\frac{1}{\hat{\alpha}}\ln{\frac{u}{\sqrt{\hat{\alpha}^{2}-u^{2}}
+\hat{\alpha}}}$ for $\epsilon=-1$. The upper branch for $\epsilon=1$ and the lower branch
for $\epsilon=-1$, $0<\lambda_{\ast}\xi^{2}<3$ involve the centre $r=0$ and the curvature
invariants diverge at $r=0$. The lower branch for $\epsilon=1$, $\hat{\alpha}<u<
\sqrt{\hat{\alpha}^{2}+1}$ and the upper brance for $\epsilon=-1$ extend from a non-zero minimum
radius outwards and the curvature invariants diverge at the minimum radius.
{\it{The only significant branch is
the lower one for $\epsilon=-1$, $\lambda_{\ast}\xi^{2}>3$ which is free of
divergencies in the curvature invariants.}} Namely, this solution ranges from a minimum
non-zero radius for $u=\sqrt{\hat{\alpha}^{2}-1}$ to a finite maximum radius for
$u=\hat{\alpha}$ and at the minimum radius it is seen from (\ref{aoc}) that the Ricci scalar
is finite (the same is true for the other curvature invariants).

\section{The geodesic equation}

Let us first write the metric (\ref{wel}) in the form
\begin{equation}
ds^{2}=-\frac{\epsilon\zeta P(\textmd{v})}{(\textmd{v}\!-\!\frac{\epsilon}{3c})^{2}}
e^{\mp 2I(\textmd{v})}dt^{2}
+\frac{\epsilon (\textmd{v}\!-\!\frac{\epsilon}{3c})^{2}}
{\big[1\!\pm\!\sqrt{P(\textmd{v})}\,\big]^{2}}
\,dr^{2}+r^{2}(d\theta^{2}+\sin^{2}{\!\theta}\,d\phi^{2})\,,
\label{wra}
\end{equation}
where $\textmd{v}$ is supposed to be implicitly solved in terms of $r$ from (\ref{afa}).
This metric arises because the relation of $d\textmd{v}$ and $dr$ is
\begin{equation}
\Big(\frac{d\textmd{v}}{dr}\Big)^{\!2}=\frac{P(\textmd{v})e^{\pm 2 I(\textmd{v})}}
{r_{0}^{2}\,[1\!\pm\!\sqrt{P(\textmd{v})}]^{2}}\Big(\textmd{v}\!-\!\frac{\epsilon}{3c}\Big)^{\!4}\,.
\label{afl}
\end{equation}
To find the timelike or null geodesics in the above spacetime, we orient as usually
the coordinate system so that the geodesic lies in the equatorial plane $\theta=\frac{\pi}{2}$.
Due to the presence of the Killing vectors $\frac{\partial}{\partial t}$,
$\frac{\partial}{\partial\phi}$, the energy $E$ and the
angular momentum $J$ of the particle are conserved, $E=-g_{tt}\frac{dt}{d\lambda}$,
$J=g_{\phi\phi}\frac{d\phi}{d\lambda}$, where $\lambda$ is an affine parameter.
The magnitude of the 4-vector of energy-momentum is given by the rest mass $\mu$ of the particle
($\mu=0$ for massless particles), so $g_{\mu\nu}\frac{dx^{\mu}}{d\lambda}\frac{dx^{\nu}}{d\lambda}
+\mu^{2}=0$ and therefore
\begin{equation}
|g_{tt}g_{rr}|\Big(\frac{dr}{d\lambda}\Big)^{2}=E^{2}-\epsilon|g_{tt}|\Big(\mu^{2}\!+\!
\frac{J^{2}}{r^{2}}\Big)\,.
\label{qwk}
\end{equation}

For massive particles we get
\begin{equation}
|g_{tt}g_{rr}|\Big(\frac{dr}{d\tau}\Big)^{2}=\tilde{E}^{2}-\epsilon|g_{tt}|\Big(1\!+\!
\frac{\tilde{J}^{2}}{r^{2}}\Big)\,,
\label{qwk}
\end{equation}
where $\tau=\lambda\mu$ is the proper time, $\tilde{E}=\frac{E}{\mu}$
the energy per unit rest mass and $\tilde{J}=\frac{J}{\mu}$ the angular momentum per
unit rest mass. We consider for simplicity radial orbits, so $J=0$ and we obtain
\begin{equation}
\Big(\frac{dr}{d\tau}\Big)^{2}=\big[1\!\pm\!\sqrt{P(\textmd{v})}\,\big]^{2}
\Big[\frac{\tilde{E}^{2}e^{\pm 2I(\textmd{v})}}{\zeta P(\textmd{v})}
\!-\!\frac{\epsilon}{(\textmd{v}\!-\!\frac{\epsilon}{3c})^{2}}\Big]\,.
\label{vfw}
\end{equation}
For $\epsilon=-1$, equation (\ref{vfw}) does not set any additional constraint on the allowed
range of $\textmd{v}$ (and accordingly of $r$) that the particle can acquire.
However, for $\epsilon=1$, the right hand side of (\ref{vfw}) should be non-negative,
that means
\begin{equation}
\frac{\zeta P(\textmd{v}) e^{\mp 2I(\textmd{v})}}{(\textmd{v}\!-\!\frac{1}{3c})^{2}}\leq
\tilde{E}^{2}\,.
\label{wkj}
\end{equation}

As we explained in the previous section, for the upper branch of case (i) with $\epsilon=1$
the solution extends from the origin $r=0$ for $\textmd{v}\rightarrow +\infty$ to a finite
radius for $\textmd{v}=\rho_{0}$. For this solution, the inequality (\ref{wkj}) means that
for any $\tilde{E}$ the allowed for the massive particle range of $\textmd{v}$ is from a minimum
value of $\textmd{v}$ up to a finite value $\textmd{v}=\textmd{v}_{f}$ which
saturates the inequality. This means that for the particle, $\textmd{v}$ cannot approach
infinity, and thus $r$ cannot approach the origin. The minimum distance $D_{f}$ from the
origin that the particle reaches is $D_{f}=r_{0}[e^{-I(\textmd{v}_{f})}\!-\!e^{-I_{\infty}}]$.
Although the curvature invariants diverge at the origin, however the origin is inaccessible by
any timelike particle.
We can also find the proper time $\tau_{\text{tot}}$ needed by an infalling timelike particle
starting from an initial parameter $\textmd{v}=\textmd{v}_{i}$ in order to reach the minimum
distance
\begin{equation}
\tau_{\text{tot}}=r_{0}\int_{\textmd{v}_{i}}^{\textmd{v}_{f}}\frac{e^{-I(\textmd{v})}d\textmd{v}}
{(\textmd{v}\!-\!\frac{1}{3c})\sqrt{\frac{\tilde{E}^{2}}{\zeta}
(\textmd{v}\!-\!\frac{1}{3c})^{2}e^{2I(\textmd{v})}\!-\!P(\textmd{v})}}\,.
\label{nkf}
\end{equation}
By definition, the upper value $\textmd{v}_{f}$ of the integration vanishes the
square root in the denominator of the integral (\ref{nkf}). However, this pole is seen to be
integrable, therefore the time $\tau_{\text{tot}}$ needed from the particle to reach the minimum
distance is finite. This means that the particle bounces at the minimum distance and
afterwards it moves outwards. Thus, the energy-dependent cosmological constant (along
with the other terms needed to make the equations consistent) acts repulsively
and is able to inhibit an infalling mass, while a simple cosmological constant is unable
to deter the collapse of the particle. {\footnote{Note that the upper branches of cases (ii), (iv)
with $\epsilon=-1$, which also extend from the origin to a finite radius and have divergent
curvature invariants at the origin, are singular since any timelike geodesic can be seen that it
needs a finite proper time to reach the origin.}} Since the particle is reflected, its geodesic
is extendible (at least in a region of its minimum approach), so the space is what is called
t-complete, i.e. it is geodesically complete for the timelike geodesics.
\newline
In this spacetime however, a radially infalling massless particle reaches the origin $r=0$ in a
finite affine parameter. Indeed, we have
\begin{equation}
\Big(\frac{dr}{d\lambda}\Big)^{\!2}=\frac{E^{2}}{\zeta}
\frac{e^{2I(\textmd{v})}}{P(\textmd{v})}\big[1\!+\!\sqrt{P(\textmd{v})}\big]^{2}
\label{ewk}
\end{equation}
and approaching the origin the radial velocity $\frac{dr}{d\lambda}$ becomes constant.
Therefore, the total affine parameter of the null geodesic
\begin{equation}
\lambda_{\text{tot}}=\frac{\sqrt{\zeta}\,r_{0}}{E}\int_{\rho_{0}}^{+\infty}
\!\!\frac{e^{-2I(\textmd{v})}}{(\textmd{v}\!-\!\frac{1}{3c})^{2}}d\textmd{v}
\label{wek}
\end{equation}
is found to be finite. Therefore, the spacetime is geodesically incomplete for the null geodesics.

For the special case $c=0$, we have seen that there are two solutions which extend from
the origin $r=0$ to a finite radius and both solutions have divergent curvature invariants at the
origin. The first refers to the upper branch of $\epsilon=1$ and the origin corresponds
to $u\rightarrow +\infty$, while the second refers to the lower branch of $\epsilon=-1$,
$0<\alpha<3$ and $r=0$ is achieved for $u=0$. For the first solution
a timelike geodesic with $\tilde{E}>\sqrt{\vartheta}\,e^{\mp\frac{\pi}{2\hat{\alpha}}}$ can be
seen that reaches the origin in finite proper time. For the second solution any
timelike geodesic reaches the origin in finite proper time. Thus, since the solutions are
geodesically incomplete and cannot be extended at the divergent origin, they are singular.
Note however that for the first solution the radial velocity $\frac{dr}{d\tau}$ tends to a constant
as the particle approaches the origin, therefore the radial acceleration is zero.

\section{Discussion and Conclusions}

General branches of new spherically symmetric solutions have been obtained in the context of
Quantum Einstein Gravity. The source is a ``cosmological constant'' field that scales at high
energies as the asymptotic safety scenario suggests, i.e.
$\Lambda(k)\propto k^2$. This scaling is also the unique one which is consistent with dimensional
analysis without the introduction of a new energy scale. The solutions solve the uniquely defined
modified Einstein equations which arise by adding appropriate covariant kinetic terms to the action in
order to ensure the conservation of energy through the Bianchi identities. The importance of
the presented solutions lies on the fact that they describe the interior small distance behaviour
of a quantum corrected black hole according to asymptotically safe gravity.
It is expected that the vacuum quantum modified spherically symmetric solutions in the UV region
is dominated by the $\Lambda$ field.

The study of the solutions revealed important new features for the spacetime structure close to the
centre of spherical symmetry. Since the signature of the metric close to these centres is not known,
we have considered both cases, the one with the ``correct'' signs for the temporal-radial metric
components $(-+++)$ and also the ``opposite'' one with signature $(+-++)$. The ``opposite'' signature
cannot be excluded since the interior Schwarzschild black hole in the same spherical coordinates
has also this signature. The physically significant solutions, i.e. those which improve or resolve
the singularity issues, are separated into two categories.

The first category consists of one solution with the ``correct'' signature which starts from the
origin $r=0$ and is accompanied by a curvature singularity at the origin. However, no timelike
geodesic can approach the origin since any infalling massive particle is reflected at some finite
distance. This is due to the repulsive force caused by the cosmological constant field along with the
other kinetic terms, and this feature is not met by a standard cosmological constant. Although the space
is timelike geodesically complete (t-completeness), still the massless particles reach the origin in
finite affine parameter.

The second category contains various solutions of both signatures which range from a finite radius
outwards and cannot be extended to involve the centre of spherical symmetry. These solutions have
finite all their curvature invariants at this minimum radius, so they can also be considered as
non-singular physically accepted geometries.

The presented general solutions which describe the interior of a quantum corrected black hole are
valid up to the length where the $G$ field starts to become significant. Thus, at that length one
should join the presented solutions with the quantum corrected one that includes both fields $\Lambda$
and $G$, which consequently for even larger length scales should turn into the the classical
Schwarzschild black hole. The present work could also be useful in other directions that utilize
quantum corrected black holes to explain baryogenesis \cite{Aliferis:2014ofa}, stability issues
\cite{Abdalla:2006qj}, or dark energy \cite{Kofinas:2011pq}.




\begin{thebibliography}{99}

\bibitem{tHooft:1974bx}
G.~'t~Hooft and M.~J.~G. Veltman,
\newblock Annales Poincar{\'e} Phys.\ Theor.\ A {\bf 20} (1974) 69.

\bibitem{Goroff:1985sz}
M.~H. Goroff and A.~Sagnotti,
\newblock Phys.\ Lett.\ B {\bf 160} (1985) 81.

\bibitem{vandeVen:1991gw}
A.~E.~M. van~de Ven,
\newblock Nucl.\ Phys.\ B {\bf 378} (1992) 309.

\bibitem{livrev}
 M.~Niedermaier and M.~Reuter, Living Reviews in Relativity {\bf 9} (2006) 5;
 M.~Reuter and F.~Saueressig, in {\it Geometric and Topological Methods for Quantum Field Theory},
 H.~Ocampo, S.~Paycha and
 A.~Vargas (Eds.), Cambridge Univ.\ Press, Cambridge, 2010, hep-th/0708.1317;
 R.~Percacci, in \textit{Approaches to Quantum Gravity: Towards a New Understanding of Space, Time and Matter}, D. Oriti (Ed.), Cambridge University Press, Cambridge, 2009,
  hep-th/0709.3851.

\bibitem{oliverbook}
O.~Lauscher and M.~Reuter in \textit{Quantum Gravity}, B.~Fauser,
J.~Tolksdorf and E.~Zeidler (Eds.), Birkh\"auser, Basel, 2007, hep-th/0511260.

\bibitem{reviews}
M. Reuter and F. Saueressig, New J. Phys. {\bf 14} (2012) 055022, hep-th/1202.2274;
B. Koch and F. Saueressig, Int. J. Mod. Phys. A {\bf 29}, no. 8, 1430011 (2014), hep-th/1401.4452;
A. Bonanno, PoS CLAQG08, 008 (2011), hep-th/0911.2727.


\bibitem{wein}
S.~Weinberg
in \textit{General Relativity, an Einstein Centenary Survey},
S. W.~Hawking and W.~Israel (Eds.),
Cambridge University Press, 1979;
S.~Weinberg, hep-th/9702027.

\bibitem{Weinproc1}
S.~Weinberg, hep-th/0903.0568; S.~Weinberg, PoS C {D09} (2009) 001, hep-th/0908.1964.

\bibitem{mr}
M.~Reuter,
Phys.\ Rev.\ D {\bf 57} (1998) 971, hep-th/9605030.
%
\bibitem{percadou}
D.~Dou and R.~Percacci,
Class.\ Quant.\ Grav.\ {\bf 15} (1998) 3449, hep-th/9707239.
%
\bibitem{oliver1}
O.~Lauscher and M.~Reuter,
Phys.\ Rev.\ D {\bf 65} (2002) 025013, hep-th/0108040.
%
\bibitem{frank1}
M.~Reuter and F.~Saueressig,
Phys.\ Rev.\ D {\bf 65} (2002) 065016, hep-th/0110054.
%
\bibitem{oliver2}
O.~Lauscher and M.~Reuter, Phys.\ Rev.\ D {\bf 66} (2002) 025026, hep-th/0205062.
%
\bibitem{oliver3}
O.~Lauscher and M.~Reuter,
Class.\ Quant.\ Grav.\ {\bf 19} (2002) 483, hep-th/0110021.
%
\bibitem{oliver4}
O.~Lauscher and M.~Reuter,
Int.\ J.\ Mod.\ Phys.\ A {\bf 17} (2002) 993, hep-th/0112089.
%
\bibitem{souma}
W.~Souma,
Prog.\ Theor.\ Phys.\ {\bf 102} (1999) 181, hep-th/9907027.
%
\bibitem{frank2}
M.~Reuter and F.~Saueressig,
Phys.\ Rev.\ D {\bf 66} (2002) 125001, hep-th/0206145;
M.~Reuter and F.~Saueressig, Fortschr.\ Phys.\ 52 (2004) 650, hep-th/0311056.

\bibitem{prop}
A.~Bonanno and M.~Reuter, JHEP\ {\bf 02} (2005) 035, hep-th/0410191.

\bibitem{perper1}
R.~Percacci and D.~Perini,
Phys.\ Rev.\ D {\bf 67} (2003) 081503, hep-th/0207033;
Phys.\ Rev.\ D {\bf 68} (2003) 044018, hep-th/0304222;
Class.\ Quant.\ Grav.\ {\bf 21} (2004) 5035, hep-th/0401071.

\bibitem{codello}
A.~Codello and R.~Percacci, Phys.\ Rev.\ Lett.\  {\bf 97} (2006) 221301, hep-th/0607128.
%

\bibitem{litimgrav}
D.~Litim, Phys.\ Rev.\ Lett.\ {\bf 92} (2004) 201301, hep-th/0312114. \\
P.~Fischer and D.~Litim, Phys.\ Lett.\  B {\bf 638} (2006) 497, hep-th/0602203.

%
\bibitem{essential}
R.~Percacci and D.~Perini,
Class.\ Quant.\ Grav.\ {\bf 21} (2004) 5035, hep-th/0401071.
%
\bibitem{r6}
A.~Codello, R.~Percacci and C.~Rahmede, Int.\ J.\ Mod.\ Phys.\  A {\bf 23} (2008) 143, hep-th/0705.1769.

\bibitem{MS1}
  P.~F.~Machado and F.~Saueressig,
Phys.\ Rev.\ D {\bf 77} (2008) 124045, hep-th/0712.0445.

\bibitem{Codello:2008vh}
  A.~Codello, R.~Percacci and C.~Rahmede,
  Annals Phys.\ {\bf 324} (2009) 414, hep-th/0805.2909.
%

\bibitem{creh1}
M.~Reuter and H.~Weyer, Phys.\ Rev.\ D {\bf 79} (2009) 105005, hep-th/0801.3287;
M.~Reuter and H.~Weyer, Gen.\ Rel.\ Grav. {\bf 41} (2009) 983, hep-th/0903.2971.
%
\bibitem{creh2}
M.~Reuter and H.~Weyer, Phys.\ Rev.\  D {\bf 80} (2009) 025001, hep-th/0804.1475.
%

\bibitem{creh3}
  P.~F.~Machado and R.~Percacci,
  Phys.\ Rev.\  D {\bf 80} (2009) 024020, hep-th/0904.2510.

\bibitem{JE1}
  J.~E.~Daum and M.~Reuter, Adv.\ Sci.\ Lett.\ {\bf 2} (2009) 255, hep-th/0806.3907.
%
\bibitem{HD1}
  D.~Benedetti, P.~F.~Machado and F.~Saueressig, Mod.\ Phys.\ Lett.\ A {\bf 24} (2009) 2233,
  hep-th/0901.2984.

\bibitem{HD2}
  D.~Benedetti, P.~F.~Machado and F.~Saueressig, Nucl.\ Phys.\ B {\bf 824} (2010) 168,
  arXiv:0902.4630; D.~Benedetti, P.~F.~Machado and F.~Saueressig, hep-th/0909.3265.



\bibitem{JEUM}
    J.-E.~Daum, U.~Harst and M.~Reuter, JHEP {\bf 01} (2010) 084, hep-th/0910.4938;
    U.~Harst and M.~Reuter, JHEP {\bf 05} (2011) 119, hep-th/1101.6007.

\bibitem{Rahmede:2011zz}
  C.~Rahmede,
  PoS CLAQG  {\bf 08} (2011) 011.

\bibitem{Benedetti:2010nr}
  D.~Benedetti, K.~Groh, P.~F.~Machado and F.~Saueressig,
  JHEP {\bf 06} (2011) 079,
  hep-th/1012.3081.

\bibitem{Bonanno:2001xi}
  A.~Bonanno and M.~Reuter,
  Phys.\ Rev.\ D {\bf 65} (2002) 043508,
  hep-th/0106133.

\bibitem{Bonanno:2001hi}
  A.~Bonanno and M.~Reuter,
  Phys.\ Lett.\ B {\bf 527} (2002) 9,
  astro-ph/0106468.

\bibitem{Reuter:2012xf}
M.\ Reuter and F.\ Saueressig,   {\it Lect.\ Notes Phys.\ } {\bf 863} (2013) 185,
 hep-th/1205.5431.


\bibitem{Bonanno:1998ye}
  A.~Bonanno and M.~Reuter,
  Phys.\ Rev.\  D {\bf 60} (1999) 084011,
  gr-qc/9811026.

\bibitem{Bonanno:2000ep}
  A.~Bonanno and M.~Reuter,
  Phys.\ Rev.\  D {\bf 62} (2000) 043008,
  hep-th/0002196.

\bibitem{Bonanno:2006eu}
  A.~Bonanno and M.~Reuter,
  Phys.\ Rev.\ D {\bf 73} (2006) 083005,
  hep-th/0602159.


\bibitem{Reuter:2006rg}
  M.~Reuter and E.~Tuiran,
  hep-th/0612037.

\bibitem{Reuter:2010xb}
  M.~Reuter and E.~Tuiran,
  Phys.\ Rev.\ D {\bf 83} (2011) 044041,
  hep-th/1009.3528.


\bibitem{Falls:2012nd}
  K.~Falls and D.~F.~Litim,
  Phys.\ Rev.\ D {\bf 89} (2014) 084002, gr-qc/1212.1821.


\bibitem{Cai:2010zh}
  Y.-F.~Cai and D.~A.~Easson,
  JCAP {\bf 1009} (2010) 002, hep-th/1007.1317.

\bibitem{Becker:2012js}
  D.~Becker and M.~Reuter,
  JHEP {\bf 1207} (2012) 172, hep-th/1205.3583.

\bibitem{Becker:2012jx}
  D.~Becker and M.~Reuter,
  hep-th/1212.4274.

\bibitem{Emoto:2005te}
  H.~Emoto,
  hep-th/0511075.

\bibitem{Ward:2006vw}
  B.~F.~L.~Ward,
  Acta Phys.\ Polon.\ B {\bf 37} (2006) 1967,
  hep-ph/0605054.

\bibitem{Falls:2010he}
  K.~Falls, D.~F.~Litim and A.~Raghuraman,
  Int.\ J.\ Mod.\ Phys.\ A {\bf 27} (2012) 1250019,
  hep-th/1002.0260.


\bibitem{Basu:2010nf}
  S.~Basu and D.~Mattingly,
  Phys.\ Rev.\ D {\bf 82} (2010) 124017;
  S.~Basu and D.~Mattingly, Phys. Rev. D 82 (2010) 124017, hep-th/1006.0718.

    \bibitem{Casadio:2010fw}
  R.~Casadio, S.~D.~H.~Hsu and B.~Mirza,
  Phys.\ Lett.\ B {\bf 695} (2011) 317;
  R.~Casadio, S.~D.~H.~Hsu and B.~Mirza, Phys. Lett. B {\bf 695} (2011) 317, gr-qc/1008.2768.


\bibitem{Lauscher:2005qz}
  O.~Lauscher and M.~Reuter,
  JHEP {\bf 0510} (2005) 050,
  hep-th/0508202.


\bibitem{Reuter:2011ah}
  M.~Reuter and F.~Saueressig,
  JHEP {\bf 1112} (2011) 012,
  hep-th/1110.5224.

\bibitem{Rechenberger:2012pm}
  S.~Rechenberger and F.~Saueressig,
  Phys.\ Rev.\ D {\bf 86} (2012) 024018,
  hep-th/1206.0657.

\bibitem{Calcagni:2013vsa}
  G.~Calcagni, A.~Eichhorn and F.~Saueressig,
  Phys.\ Rev.\ D {\bf 87}, 124028 (2013),
  hep-th/1304.7247.

\bibitem{avoid}
R. Torres, Phys. Lett. B {\bf 733} (2014) 21, gr-qc/1404.7655;
R. Torres and F. Fayos, Phys. Lett. B {\bf 733} (2014) 169, gr-qc/1405.7922;
R. Casadio, S. D. H. Hsu, B. Mirza, Phys. Lett. B {\bf 695} (2011) 317, gr-qc/1008.2768.

\bibitem{Koch:2013owa}
  B.~Koch and F.~Saueressig,
  Class.\ Quant.\ Grav.\   {\bf 31} (2013) 015006,
  hep-th/1306.1546.

\bibitem{Koch:2013rwa}
  B.~Koch, C.~Contreras, P.~Rioseco and F.~Saueressig,
  hep-th/1311.1121.

\bibitem{bianchi}
M. Reuter, H. Weyer, Phys. Rev. D {\bf 69} (2004) 104022, hep-th/0311196;
M. Reuter, H. Weyer, Phys. Rev. D {\bf 70} (2004) 124028, hep-th/0410117.


\bibitem{GradR} I. S. Gradshteyn and I. M. Ryzhik, {\it{Table of Integrals, Series and Products}},
7th Ed., Academic Press, 2007.

\bibitem{Prud} A. P. Prudnikov, Yu. A. Brychkov and O. I. Marichev, {\it{Integrals and Series}},
Vol 1, Overseas Pub. Ass., 1986.

\bibitem{AbraS} M. Abramowitz and I. A. Stegun, {\it{Handbook of Mathematical Functions with
formulas, graphs and mathematical tables}}, Dover Pub., New York, 1972.

\bibitem{Aliferis:2014ofa}
  G.~Aliferis, G.~Kofinas and V.~Zarikas,
  Phys.\ Rev.\ D {\bf 91} (2015) 4, 045002, hep-ph/1406.6215.


\bibitem{Abdalla:2006qj}
  E.~Abdalla, B.~Cuadros-Melgar, A.~B.~Pavan and C.~Molina,
  Nucl.\ Phys.\ B {\bf 752} (2006) 40, gr-qc/0604033;
  B.~Cuadros-Melgar, J.~de Oliveira and C.~E.~Pellicer,
  Phys.\ Rev.\ D {\bf 85} (2012) 024014, hep-th/1110.4856.


\bibitem{Kofinas:2011pq}
  G.~Kofinas and V.~Zarikas,
  Eur.\ Phys.\ J.\ C {\bf 73} (2013) 4, 2379, hep-th/1107.2602.



\end{thebibliography}
\end{document}